\documentclass{pasa}%

\title[The $\gamma$-ray pulsar J0633+0632 in X-rays]{The $\gamma$-ray pulsar J0633+0632 in X-rays}
\author[Danilenko et al.]{Andrey Danilenko$^1$, Peter Shternin$^{1}$, Anna Karpova$^{1,2}$, Dima Zyuzin$^1$ \and Yuriy Shibanov$^{1,2}$ 
\\
\affil{$^1$Ioffe Institute, Politekhnicheskaya
 26, St.~Petersburg, 194021, Russia}%
\affil{$^2$Peter The Great St. Petersburg Polytechnic University,
Politekhnicheskaya 29, St.~Petersburg, 195251, Russia}
\affil{Email: danila@astro.ioffe.ru}
}%

\jid{PASA}
\doi{10.1017/pas.\the\year.xxx}
\jyear{\the\year}

\usepackage{natbib}
\bibpunct{(}{)}{;}{a}{}{,}

\usepackage{pdflscape}
\usepackage{amsmath}

\def\asec{\ifmmode ^{\prime\prime}\else$^{\prime\prime}$\fi}

\def\it{\sl} 
\def\degs{\ifmmode ^{\circ}\else$^{\circ}$\fi}
\def\amin{\ifmmode ^{\prime}\else$^{\prime}$\fi}
\def\asec{\ifmmode ^{\prime\prime}\else$^{\prime\prime}$\fi}
\def\fdg{\hbox{$.\!\!^\circ$}}          
\def\farcs{\hbox{$.\!\!^{\prime\prime}$}}  

\def\degs{\ifmmode ^{\circ}\else$^{\circ}$\fi}
\def\amin{\ifmmode ^{\prime}\else$^{\prime}$\fi}
\def\farcm{\hbox{$.\mkern-4mu^\prime$}}
\def\eqalign#1{\null\,\vcenter{\openup1\jot \m@th
   \ialign{\strut\hfil$\displaystyle{##}$&$\displaystyle{{}##}$\hfil
   \crcr#1\crcr}}\,}

\begin{document}%

\begin{abstract}
  We analysed \textit{Chandra} observations of the bright
  \textit{Fermi} pulsar J0633+0632 and found evidence of an absorption feature in
  its spectrum at $804^{+42}_{-26}$~eV (the errors here and below are at 90\% confidence) with equivalent width of $63^{+47}_{-36}$ eV.
  In addition, we analysed in detail the X-ray spectral continuum taking into account
  correlations between the interstellar absorption and the distance to
  the source. We confirm early findings by \citet{ray2011ApJS} that the spectrum contains non-thermal and
  thermal components. The latter is equally well described by the
  blackbody and magnetised atmosphere models and can be attributed
  to the emission from the bulk of the stellar surface in both cases.
  The distance to the pulsar is constrained
  in a range of 1--4~kpc from the spectral fits. We infer the
  blackbody surface temperature of
  $108^{+22}_{-14}$~eV, while for the atmosphere model, the temperature, as seen by a distant
  observer, is $53^{+12}_{-7}$~eV. In the latter case J0633+0632
  is one of the coldest middle-aged isolated neutron stars with measured temperatures.
  Finally, it powers an extended pulsar wind
  nebula whose shape suggests a high pulsar proper motion. Looking backwards the
  direction of the presumed proper motion we found
  a likely  birthplace of the pulsar -- the Rosette nebula, a 50-Myr-old active star-forming region
  located at about 1\fdg5 from the pulsar. If true, this
  constrains the distance to the pulsar in the range of
  1.2--1.8~kpc.
\end{abstract}
\maketitle%
\section{INTRODUCTION }
\label{sec:intro} Usually, X-ray spectra of isolated neutron stars
(NSs) are well described by a featureless continuum which contains
nonthermal and/or thermal component. In rare cases, however,
absorption features are detected. Understanding  origins of these
features is thought to be important for various aspects of the NS
physics. For instance, they can result from atomic transitions in
the mid-Z element NS atmospheres \citep[e.g.,][]{mori2007MNRAS}.
In this case, as in ordinary stars, it is possible to measure the
surface gravitational redshift and, hence, the stellar mass to
radius ratio. This is important for independent diagnostic of the
equation of state (EOS) of dense matter inside NSs. Absorption
features can be also identified with either proton or electron
cyclotron lines. Electron cyclotron lines in NS spectra were
predicted by \citet{gnedin74} and then discovered by
\citet{truemper78} in the Her X-1 binary system. These lines
were detected in many accreting X-ray pulsars since then,
\citep[e.g.,][]{Revnivtsev2014} allowing for direct measurements
of NS magnetic fields.

For isolated NSs, until recently, absorption features have been seen in X-ray
spectra of only a few atypical, radio-silent, pure thermally emitting
sources. This includes two low-magnetic-field central
compact objects (CCOs) in supernova remnants (SNRs), 1E 1207$-$5209
\citep{sanwal2002ApJ} and PSR J0821$-$4300
\citep{gotthelf2009ApJ}, five objects with larger fields from the
``Magnificent Seven'' family \citep[e.g.,][and references
therein]{pires2014AsAp} and one soft gamma repeater SGR 0418+5729 \citep{tiengo2013Natur}. Sole
exception is an ordinary middle-aged radio pulsar J1740+1000
\citep{kargaltsev2012Sci}.

Since the launch of the \textit{Fermi} $\gamma$-ray observatory,
several dozens of new pulsars \citep{abdo2013ApJS} have been
discovered. A substantial number of them are not seen in
radio but are identified in X-rays. The radio-quiet PSR
J0633+0632 (hereafter J0633) was discovered in a blind search for
pulsations in the \textit{Fermi}-LAT data  \citep{abdo2009}. Among
\textit{Fermi}-pulsars, J0633 is one of the brightest in X-rays,
with a flux $F_{\rm X}$~$\sim$~$10^{-13}$~erg~cm$^{-2}$~s$^{-1}$
\citep{ray2011ApJS}. A pulsar period $P$~= 297.4~ms and a rotation
frequency derivative imply a characteristic age $\tau$~= 59.2~kyr,
a spin-down luminosity $\dot{E}$~= $1.2\times10^{35}$~erg~s$^{-1}$
and a surface magnetic field $B$~= $4.9\times10^{12}$~G
\citep{abdo2013ApJS}. A distance $D$~$\sim$~1 kpc was estimated
from an empirical relation between $\gamma$-ray and spin-down
luminosities \citep{sazparkinson2010ApJ}. In addition, J0633
powers an extended pulsar wind nebula (PWN) visible in X-rays
south from the pulsar. Analyzing 20-ks \textit{Chandra}
observations, \citet{ray2011ApJS} found that the X-ray spectrum of
J0633 contains a thermal component which dominates at low energies
and a non-thermal power-law (PL) component of the NS
magnetospheric origin describing the high energy spectral tail.

Re-analysing the \textit{Chandra} data, we found a hint of an absorption feature in
the spectral fit residuals. We argue
that this feature is real in Section~\ref{sec:an_feature} and discuss
its possible nature in Section~\ref{sec:disc_feature}.
In Section~\ref{sec:an_continuum}, we analyse the X-ray spectral continuum.
We confirm findings of
\citet{ray2011ApJS} and extend their analysis by incorporating
natural constraints on the interstellar absorption and the distance to
the pulsar. We analyse in detail the thermal component and
investigate whether it can be attributed to the emission from the
entire NS surface or a substantial part of it. In such a case,
it is possible to confront inferred temperatures with
predictions of NS cooling theories; this is  done in
Section~\ref{sec:disc_cool}. In addition, a speculative birthplace
of J0633 is proposed in Section~\ref{sec:birth}.
If it is real, it provides additional independent
constraints on the distance to the pulsar.

\section{ANALYSIS OF THE X-RAY DATA}
\label{sec:an}

\begin{figure}[t]
  \includegraphics[scale=0.46]{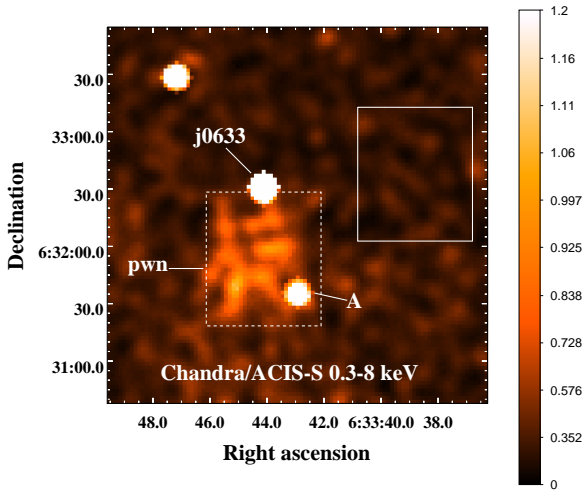}
 \caption{Field of J0633 as seen with \textit{Chandra}/ACIS-S in the 0.3--8 keV range.
 The image is binned by 4 ACIS pixels and smoothed with a 4-pixel Gaussian kernel. The pulsar
 is marked in the image and its PWN adjacent to the pulsar from south is clearly seen.
 $60\asec\times70\asec$ dashed box shows the region used to extract the PWN spectrum.
 The unrelated background source `A'
 falling in the extraction region is also marked.
 The solid rectangle with dimensions of $60\asec\times70\asec$
 shows the region used for the background extraction.
 The intensity is given in counts per pixel.}
 \label{fig:chandra}
\end{figure}

We retrieved the data\footnote{PI Roberts,
\textit{Chandra}/ACIS-S, Exp. time 20 ks, OBsID 11123} from the
\textit{Chandra} archive. Data mode was VFAINT, exposure mode was
TE and the pulsar was exposed on the ACIS-S3 chip. The
\texttt{CIAO v.4.6} \texttt{chandra\_repro} tool with
\texttt{CALDB v.4.5.9} was used to reprocess the data set. A
fragment of the \textit{Chandra} image of the pulsar field is
shown in Figure~\ref{fig:chandra}. The pulsar and its extended PWN
are clearly seen in the image. We extracted spectra of the pulsar
and the PWN in the range of 0.3--10 keV with the \texttt{CIAO}
\texttt{v.4.6} {\sl specextract} tool. For the background, we used
a region free from any sources which is shown by the solid
rectangle in Figure~\ref{fig:chandra}. The PWN spectrum was
extracted from the dashed rectangle shown in
Figure~\ref{fig:chandra} excluding the pulsar and the point-like
background object `A' which overlap with the PWN. The number
of counts for the PWN and the background in the same region were
 397 and 402, respectively. To extract the
pulsar spectrum, we used a circular aperture centred at the pulsar
position with the radius of 2\farcs5, which ensures maximal
signal-to-noise ratio. There are 332 pulsar counts ($\gtrsim$ 98\%
of the total number of pulsar counts), two counts of background
and two counts of the PWN within the aperture.

We fitted the pulsar spectrum by an absorbed sum of
power-law (PL) and thermal components using the \texttt{XSPEC}
\texttt{v.12.8.2} package \citep{arnaud1996ASPC}. To account for
the photoelectric absorption we used the \texttt{XSPEC}
PHABS model with default cross-sections \texttt{bcmc}
\citep{bcmc1992ApJ} and abundances \texttt{angr}
\citep{angr1989GeCoA}. For the thermal component we tried
blackbody (BB) and hydrogen magnetic  atmosphere models NSA
\citep{pavlov1995lns} and NSMAX \citep{ho2008ApJS}. Since
the number of the pulsar counts is small, binning the spectrum goes at
the expense of spectral resolution. Therefore, we used unbinned pulsar
spectrum in our analysis. Accordingly, we employed the
$C$-statistic \citep{cash1979ApJ} for fitting, instead of
more common $\chi^{2}$.

We performed the fitting by
a Markov chain Monte-Carlo (MCMC) sampling procedure assuming
uniform prior distribution for model parameters.
We employed the affine-invariant MCMC sampler developed by \citet{goodman2010CAMCS}
and implemented in a \texttt{python} package
\texttt{emcee} by \citet{foreman-mackey2013PASP}. For each model
we used a set of 100 MCMC walkers performing 1500 steps after
initial burning, which is large enough considering that typical
autocorrelation time \citep[see, e.g.,][for
details]{goodman2010CAMCS} was of the order of several tens (50--80).
This resulted in a set of 150000 samples in total which was enough to
reliably approximate the posterior distribution of the model parameters.
Having the sampled posterior distribution we obtained best-fit estimates and
credible intervals\footnote{The credible interval is any
continuous part of the parameter's marginal distribution
containing certain fraction of the total distribution
\citep{GelmanBook}. Here we adopt the range between the 5\% and
95\% quantiles, unless stated otherwise.} of the model parameters,
and corresponding values of $C$-statistic.

To assess the goodness of fit for each model we simulated spectra under
the model in question with parameters
drawn from the corresponding sampled posterior distribution.
Fitting these spectra
we obtained the reference distribution of the  $C$-statistic.\footnote{This procedure is close to
what the \texttt{XSPEC} goodness task does. The goodness task
uses the standard bootstrapping scheme where the best-fit estimates of the model parameters are used to simulate spectra.
The approach we employed is a more general as it incorporates
the model parameter uncertainties conditional on the current observational data.}
The $C$-statistic value obtained for the real data  was compared with this distribution.
Goodness-of-fit tests performed
in such a way showed that any one-component model, pure PL or pure thermal,
fails to describe the data (for 100\% of simulated spectra, the $C$-statistic is less than observed one).
In contrast, all two-component models mentioned above fit the data well, that is,
the observed $C$-statistic value in each case is within one standard deviation from the mean of the corresponding
reference distribution.
Similarly, we found that an absorbed PL model is consistent with the spectrum of PWN.

\subsection{Absorption feature}
\label{sec:an_feature}

\begin{figure*} 
  \setlength{\unitlength}{1mm}
  \begin{picture}(160,60)(0,0)
    \put (0.0,0.0) {\includegraphics[scale=0.4]{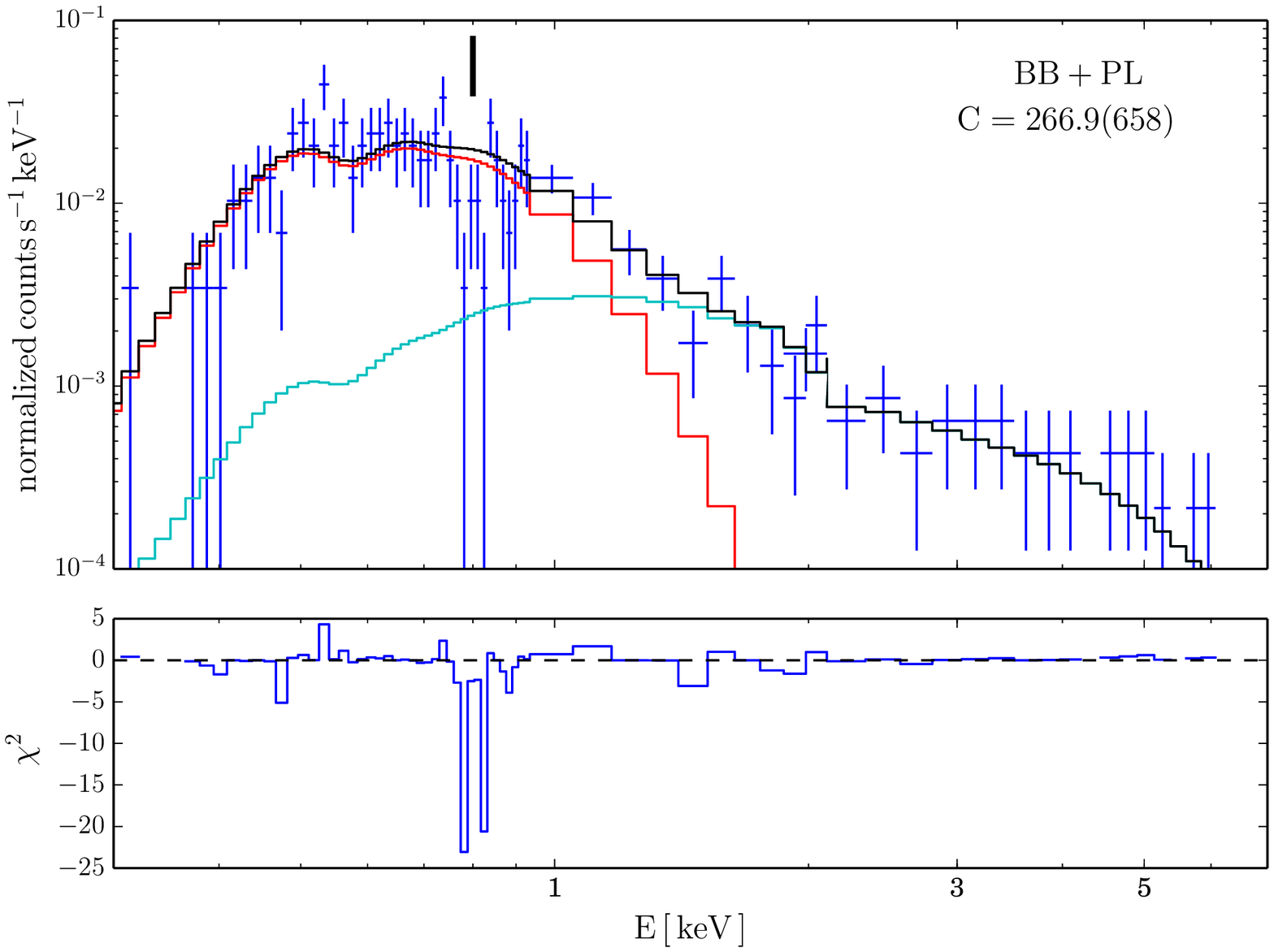}}
    \put (85.0,0.0) {\includegraphics[scale=0.4]{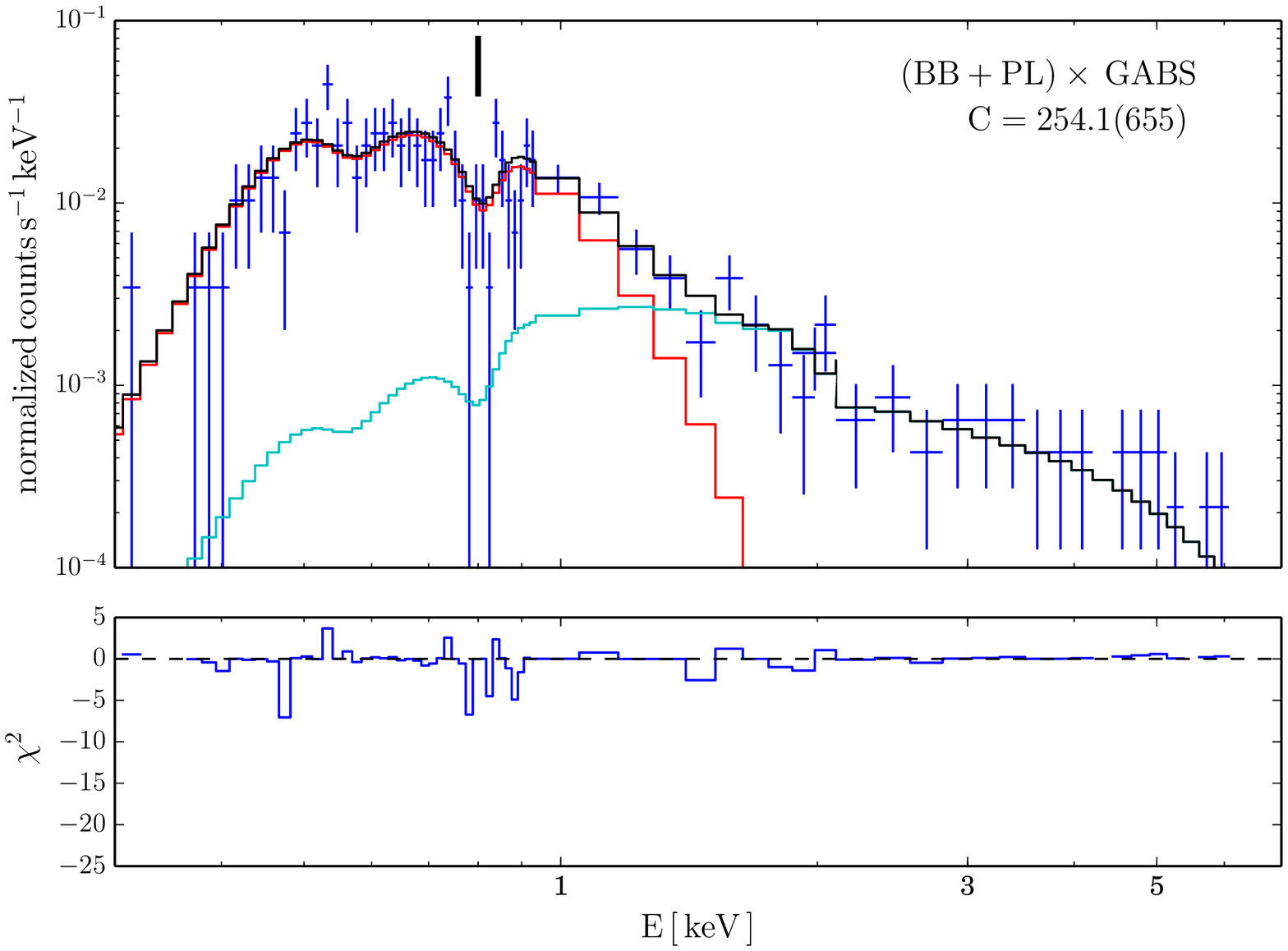}}
  \end{picture}
  \caption{\textit{Chandra}/ACIS-S spectrum of
  J0633 fitted by an absorbed BB+PL ({\it left}) and
  by an absorbed $\rm (BB+PL)\times GABS$ models ({\it right}).
  Best-fit models are shown with black solid lines in the {\it top} panels.
  Red and cyan lines show BB and PL model components, respectively.
  The fit residuals in form of $\chi^{2}$ are shown in the corresponding {\it bottom} panels.
  The absorption line position is shown by the thick bars.
  Best-fit $C$-statistic values are shown for both fitting models in the {\it top} panels with
  the values of the fit degrees of freedom given in parentheses.}
  \label{fig:spectra}
\end{figure*}

The pulsar spectrum is shown in the top-left panel of
Figure~\ref{fig:spectra} together with the best-fit
PHABS$\times$(BB+PL) model. The $C$-statistic value of 266.9 is
also presented in the plot. The corresponding fit residuals are
shown in the bottom left panel of Figure~\ref{fig:spectra}. While
the unbinned spectrum is used for fitting, the hard-energy part of
the spectrum in Figure~\ref{fig:spectra} is binned for
illustration purposes.\footnote{We keep spectrum unbinned up to
channel 64, group 8 channels into 1 bin for channels 64--128 and
16 channels into 1 bin for channels 128--1024.}

What attracted our attention, was a hint of an absorption feature
in the fit residuals at about 0.8~keV. The approximate feature
position is marked by the thick bars in the top panels of
Figure~\ref{fig:spectra}. There are at least five
consecutive channels which seem to stay apart from the best-fit
model. We thus added the Gaussian absorption (GABS) component to
the model and refitted the data. The pulsar spectrum with the new
best-fit model and corresponding fit residuals are shown in
the top-right and bottom-right panels of
Figure~\ref{fig:spectra}, respectively. The
new fit gives better $C$-statistic value of 254.1. Similar effect
was observed for all continuum
models we had tested.\footnote{This result would be the same with any model
for continuum which is smooth across the putative line region.
}

\begin{figure}[t]
  \begin{center}
   \includegraphics[scale=0.4]{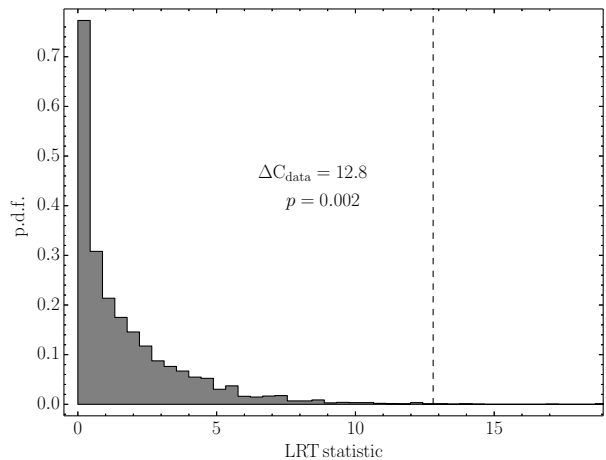}
  \end{center}
  \caption{Probability density function (p.d.f.) for the likelihood ratio test (LRT) statistic,
  that is, the difference in the $C$-statistic
  for the BB+PL and $\rm (BB+PL)\times GABS$ fits for 5000  simulated data sets.
  Vertical dashed line indicates the observed LRT statistic
  $\Delta C_{\rm data}$~= 12.8. The corresponding $p$-value is also shown, see text for details.}
 \label{fig:protassov}
\end{figure}

The difference in the $C$-statistic values between the models
with and without the line is $\Delta C=266.9-254.1=12.8$.
In order to estimate the statistical significance of the fit
improvement, we constructed the appropriate reference distribution for $\Delta C$,
or likelihood ratio test (LRT) statistic, in a manner similar to the procedure of assessing the goodness of
fit.
We simulated spectra under the model without line (the null model), drawing parameters from the
corresponding posterior distribution sampled via MCMC. We
then fitted simulated spectra with the null model and the model with line (the trial model) and computed
the corresponding $\Delta C$ or LRT statistic.
The LRT distribution basing on 5000 data sets simulated with BB+PL as a null model is
shown in Figure~\ref{fig:protassov},
where $\Delta C$~= 12.8 obtained for the data is shown by the vertical dashed line.
It is seen,
that only for 9 out of 5000 simulations the improvement in
$C$-statistic was greater than 12.8. This means that such an
improvement can hardly happen by chance if the null model is the
true one. This statement can be quantified by the
posterior predictive $p$-value, that is, a fraction of simulations
with the LRT larger than the observed one. In our case $p$-value is
0.002 which favours the absorption line presence. Similar analysis performed
with other continuum models resulted in $p$-values of the same
order of magnitude.
This method is known as
a method of posterior predictive $p$-values, the Bayesian model checking
approach recommended by \citet{protassov2002ApJ}.
In particular, \citet{protassov2002ApJ} argue that this method is superior to the more common $F$-test
in assessing the presence of additional model component.
We refer the reader to, e.g.,  \citet{GelmanBook}, for
a textbook description of the method.

\begin{figure*}[t]
  \begin{center}
   \includegraphics[scale=0.45]{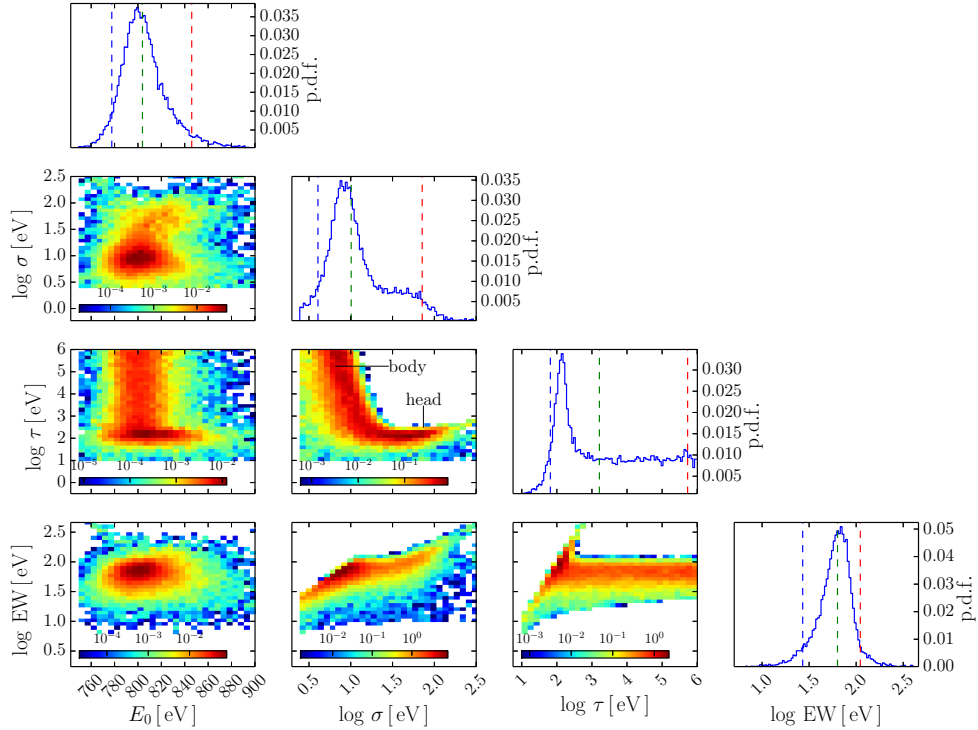}
  \end{center}
  \caption{1D and 2D marginal posterior probability distributions
  for the line parameters (line central energy $E_0$, width $\sigma$,
  depth $\tau$ and equivalent width $\rm EW$) in
  the $\rm (BB+PL)\times GABS$ model. The 5\%, 50\% and 95\% quantiles are
  shown with vertical dashed lines in the 1D distributions.}
 \label{fig:line_marginal}
\end{figure*}

\begin{table}[ht]
\caption{Median values of the absorption feature parameters with BB+PL as a
continuum model. }\label{tab:line_par}
\begin{center}
\begin{tabular}{*{4}{c}}
\hline \hline \\
$E_0$ (eV)             & $\sigma$ (eV)    & $\tau$ (eV)   & $\rm EW$ (eV) \\ \hline \\
$804^{+42}_{-26}$      & $\lesssim 285$   & $\gtrsim 10$  &  $63^{+47}_{-36}$ \\
\hline
\end{tabular}
\end{center}

\tabnote{90\% credible intervals for the line centre $E_0$ and equivalent width $\rm EW$ are
given, while 99.9\% limits for the line depth $\tau$ and the width
$\sigma$ are presented.}
\end{table}

The best-fit spectral line parameters and their uncertainties extracted from
MCMC are presented in Table~\ref{tab:line_par}. Here we use the Gaussian line model which  contains
three parameters and is given by the following expression
\begin{equation}\label{eq:gabs}
 {\rm GABS}(E)=\exp\left( - \frac{\tau}{\sqrt{2\pi}\sigma} \, {\rm
 e}^{-\frac{(E-E_0)^2}{2\sigma^2}}\right),
\end{equation}
where $E$ is the photon energy, $E_{0}$ is the line centre,
$\sigma$
is related to the full width at half maximum (FWHM) of the line as FWHM~$\approx$
2.35$\sigma$, and parameter $\tau$ regulates the line depth.
Then the optical depth at line centre is $\tau/(\sqrt{2\pi}\sigma)$.
A more direct measure of a line strength is the equivalent width (EW) which is defined by the following expression
\begin{equation}\label{eq:ew}
  \mathrm{EW} = \int\limits_{-\infty}^{+\infty}
  \left(1-\mathrm{GABS}(E)\right)\, \mathrm{d}E
\end{equation}
The main advantage of EW is that it
is weakly dependent on the particular shape of the spectral feature.
For the Gaussian line in the optically thin regime ($\tau/\sigma$~$\ll$ 1), EW~$\approx$ $\tau$.

In  Figure~\ref{fig:line_marginal}, we show one- and
two-dimensional (1D and 2D) marginal posterior distributions for
the line parameters. As seen, $E_{0}$ is well constrained around
0.8 keV. The median value of $E_{0}$ along with the 90\% credible
interval is presented in Table~\ref{tab:line_par}. Unfortunately,
the situation is different for the line width and depth.
Figure~\ref{fig:line_marginal} reveals a bimodal, worm-like, 2D
posterior distribution for $\sigma$ and $\tau$ parameters. This
bi-modality is also seen in the 1D distributions for these
parameters. The ``worm'' head and body correspond to different
types of the absorption line. The worm-body mode corresponds to a
strong saturated line with the width smaller than the
\textit{Chandra}/ACIS-S spectral resolution (FWHM $\sim$ 100
eV\footnote{see
http://cxc.harvard.edu/proposer/POG/html/chap6.html}). The fit
quality is then determined by the wings of the line, which
results, as can be shown analytically, in a strong degeneracy
between the line width and depth giving the long worm-body valley
in the likelihood distribution. On the other hand, the worm head
corresponds to a broader and weaker line. With the present data we
cannot discriminate between these possibilities. Therefore, only
the upper limit on $\sigma$ and lower limit on $\tau$ are given in
Table~\ref{tab:line_par}. At the same time, EW is well constrained
as seen from Figure~\ref{fig:line_marginal} and
Table~\ref{tab:line_par} which can be regarded as the most
straightforward argument in favour of the line.

Note that the above results are almost independent on the
particular continuum model used to fit the pulsar spectrum. In
addition, they remain qualitatively the same if models other than
GABS are used to fit the absorption feature, for instance the
models for a cyclotron absorption line (CYCLABS in \texttt{XSPEC})
or an ionization edge (EDGE in \texttt{XSPEC}).

\begin{figure}[t]
 \begin{center}
  \includegraphics[scale=0.42]{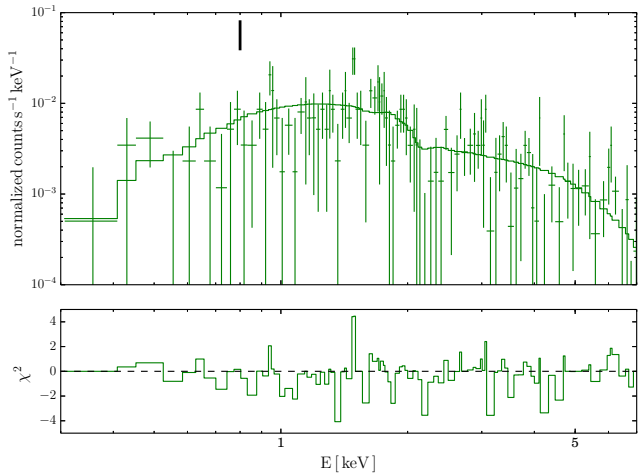}
 \end{center}
  \caption{\textit{Chandra}/ACIS-S spectrum of
  the J0633 PWN fitted by an absorbed PL. The spectrum is binned to ensure $\geq$~1 counts per bin.
  The absorption line position is shown by the thick bar.
  }
  \label{fig:spectra_pwn}
\end{figure}

The Bayesian analysis shows that the chances are low that the
absorption feature is caused by Poisson fluctuations of the data
counts. It may be an instrumental artifact though. It would
thereby be  appropriate to examine if there are similar features
in spectra of other sources in the \textit{Chandra}/ACIS-S field
of view. However, all other point sources are substantially dimmer
than the pulsar, showing no more than several tens of counts, and
analysis of their spectra is not conclusive. We also examined the
spectrum of the PWN (see Figure~\ref{fig:spectra_pwn}).
Unfortunately, the PWN spectrum is more noisy at the soft energies
than the pulsar spectrum due to much higher background. There is
no line seen here, at least at the first sight. Indeed, we got
improvement in statistic of only $\Delta C$~$\approx$ 1.5
after fitting the PWN spectra with the $\rm PL\times GABS$ model
with $E_{0}$ of about 0.8 keV. The posterior-predictive $p$-value
test gives no evidence against the featureless continuum model
with $p$-value~$\approx$ 0.43. We also checked that there were no
flares during observations which could distort the spectrum.
Accordingly, the background spectrum is in agreement with the
quiescent spectrum of the diffuse soft X-ray background as seen
with \textit{Chandra}/ACIS \citep{markevitch2003ApJ}.

\subsection{X-ray continuum}
\label{sec:an_continuum}

In this section we employ the same Bayesian technique to analyse the
properties of the X-ray continuum assuming the presence of the line.
We will not explicitly indicate this
further for brevity.

Main problems arising during the analysis of the X-ray  emission
from NSs come from unknown distances $D$ and interstellar
absorption towards these objects. In order to get better
constrains on  the latter, we now fit simultaneously the pulsar
and the PWN spectra, tying the value of the hydrogen column
density $N_{\rm H}$ between the fits. Recall that the PWN spectrum
is well described by an absorbed PL, while the pulsar spectrum
contains thermal and non-thermal components. For the thermal
component, as already mentioned above, we tested simple blackbody
as well as several magnetised hydrogen atmosphere models. All
latter give similar results, therefore we selected a particular
model from NSMAX family, labeled 1260 \citep[see][]{ho2008ApJS}.
The reason for selection of this model is twofold. First, NSMAX
models account for the partial ionization in the NS atmospheres
which makes their usage more physically motivated for low
temperatures in comparison with older NSA models. Second, the 1260
model corresponds to the surface magnetic field of
$5\times10^{12}$~G which is close to the J0633 value as inferred
from $P$--$\dot{P}$ observations assuming dipole losses. Any
atmospheric model depends on the NS surface gravity. In the NSMAX
models, this is incorporated via the gravitational redshift
parameter $1+z$. In our fit we fixed $1+z=1.21$ which corresponds
to a reasonable NS model with a mass $M_{\rm NS}$~= 1.4$M_\odot$
and a circumferential radius $R_{\rm NS}=13$~km. The apparent NS
radius as seen by a distant observer in that case is
$R=(1+z)R_{\rm NS}\approx 16$~km. We have checked that the
redshift parameter is not constrained by the data and does not
influence the final results. We also note that, due to effects of
general relativity, the NSMAX model can be applied only for
describing the emission coming from the entire surface of the NS,
while the blackbody model can be used to describe the emission
from any part of the stellar surface.

\begin{table*}[th]
\scriptsize
 \caption{Best-fit spectral parameters for continuum models. 
  }\label{t:x-fit}
  \begin{tabular}{l*{9}{c}}
\\ \hline \hline \multicolumn{1}{c}{} \\
Model    & $N_{\rm H}$              & $\Gamma_{\rm psr}$  & $K_{\rm psr}$          &  $T$             &  $R$ $^{a}$         &   $D$ $^{a}$        & $\Gamma_{\rm pwn}$  & $K_{\rm pwn}$         & C/d.o.f. $^{b}$  \\
         & (10$^{21}$ cm$^{-2}$) &                     & (10$^{-6}$ ph                  &  (eV)               &  (km)                  &   (kpc)               &                     & (10$^{-6}$ ph                  &   \\
         &                     &                     & keV$^{-1}$~cm$^{-2}$~s$^{-1}$) &                   &                      &                     &                     & keV$^{-1}$~cm$^{-2}$~s$^{-1}$) &         \\
\hline
\multicolumn{10}{c}{}\\
\multicolumn{10}{c}{No prior}\\
\multicolumn{10}{c}{}\\
BB+PL    & $2.4^{+1.8}_{-1.4}$ & $1.6^{+0.6}_{-0.6}$ & $9.6^{+7.6}_{-4.7}$          & $105^{+23}_{-18}$ &  $2.4^{+5.4}_{-1.5}$  &  $7^{+12}_{-5}$  &  $1.2^{+0.3}_{-0.3}$ &  $26.7^{+12.1}_{-7.5}$  & 381.7/792  \\
\multicolumn{10}{c}{}\\
NSMAX+PL &  $2.9^{+1.8}_{-1.4}$ &  $1.4^{+0.6}_{-0.6}$ &  $6.7^{+6.2}_{-3.6}$           &  $41^{+15}_{-11}$ &  $36^{+205}_{-29}$    &  $0.53^{+1.98}_{-0.45}$ &  $1.3^{+0.4}_{-0.3}$ &  $29.6^{+12.8}_{-8.4}$       & 388.3/792  \\
\multicolumn{6}{c}{}\\
\hline
\multicolumn{10}{c}{}\\
\multicolumn{10}{c}{With prior}\\
\multicolumn{10}{c}{}\\
BB+PL    &  $2.2^{+1.3}_{-1.2}$ &  $1.6^{+0.6}_{-0.6}$ &  $9.3^{+6.6}_{-4.6}$      &   $108^{+22}_{-14}$  &  $5^{+11}_{-4}$ &  $2.1^{+2.2}_{-1.3}$ &  $1.2^{+0.3}_{-0.3}$  &  $26.1^{+9.6}_{-7.3}$    & 383.2/791 \\
\multicolumn{10}{c}{}\\
NSMAX+PL &  $1.7^{+0.6}_{-0.7}$ &  $1.2^{+0.6}_{-0.6}$ &   $4.9^{+4.9}_{-2.6}$      &  $53^{+12}_{-7}$   &  $12^{+8}_{-9}$      &  $1.3^{+1.1}_{-0.6}$ &  $1.1^{+0.2}_{-0.2}$ &  $23.3^{+5.9}_{-5.2}$         & 404.4/791\\
\multicolumn{10}{c}{}\\
\hline
\end{tabular}
\tabnote{Temperatures $T$ and emitting area radii $R$
are given as measured by a distant observer.
Redshift parameter for NSMAX models is fixed at 1.21.
$\Gamma$ and $K$ are the photon index and the normalisation of the PL component.
All errors correspond to 90\% credible intervals derived via MCMC. For models in two
last lines an informative prior which includes information on the $N_{\rm H}$--$D$ correlation is applied,
see text for details.}
\tabnote{$^a$ In the ``no prior'' case,
$R$ is given assuming $D$~= 1~kpc and $D$ is given assuming $R$~= 16~km.}
\tabnote{$^b$ The number of degrees of freedom is different from those given
in Figure~\ref{fig:spectra} because here the PWN spectrum is included.}
\end{table*}

In first two rows of Table~\ref{t:x-fit}, we show best-fit
parameters of the BB+PL and NSMAX+PL models with uncertainties
corresponding to 90\% credible intervals. The latter were inferred
from marginal Bayesian posterior distributions as described above.
These results are consistent with those of
\citet{ray2011ApJS} and we thereby confirm their findings.
The values of $C$-statistic are also shown in Table~\ref{t:x-fit}.

\begin{figure*}[t]
  \setlength{\unitlength}{1mm}
  \begin{picture}(160,190)(0,0)
    \put (0.0,95.0) {\includegraphics[scale=0.4]{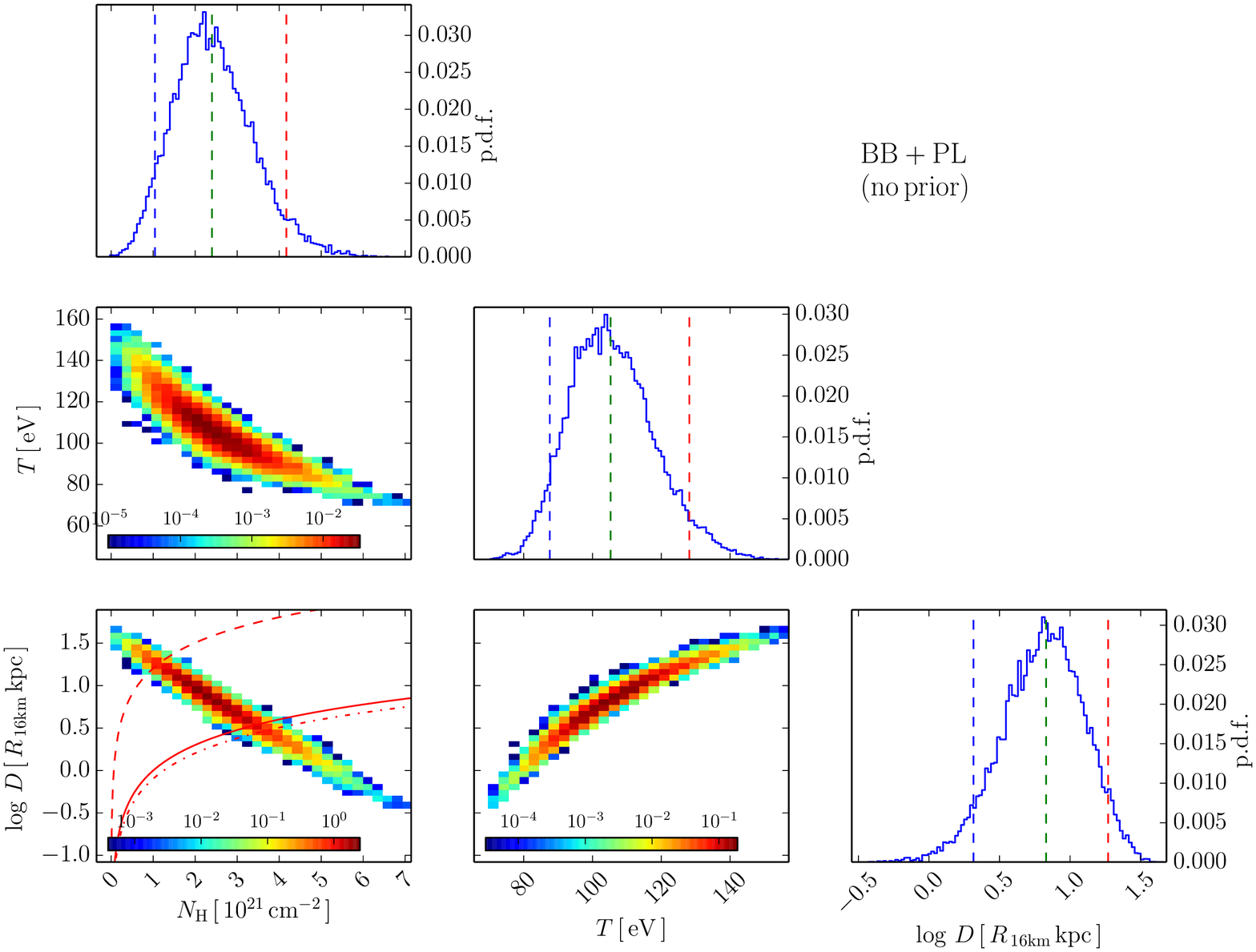}}
    \put (0.0,0.0) {\includegraphics[scale=0.4]{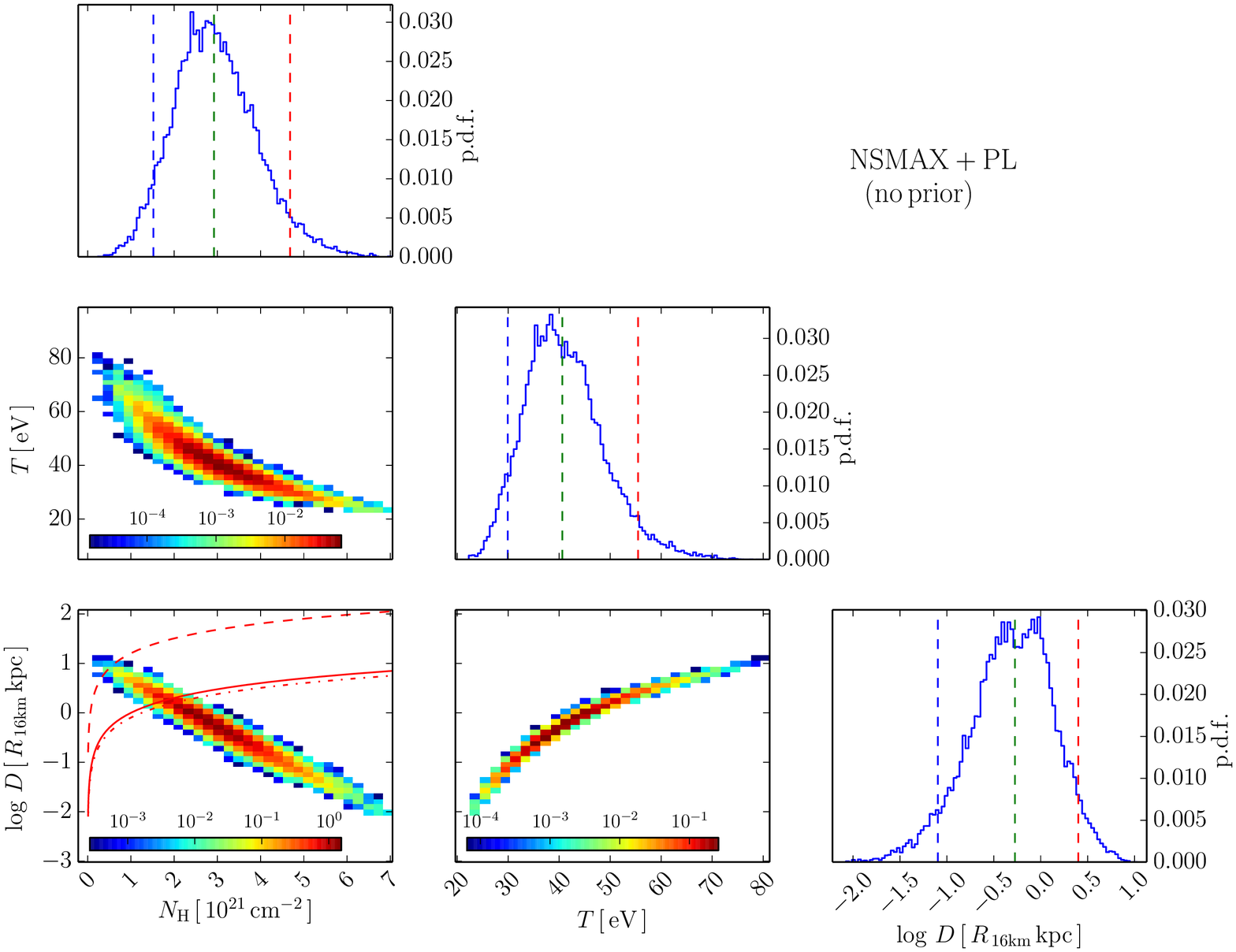}}
  \end{picture}
  \caption{{\sl Top:} 1D and 2D marginal posterior distributions for $N_{\rm H}$, $T$ and $D$
  in the BB+PL model without the prior. Distance $D$ is
  shown in units of $R_{16 \rm km}$. The solid, dashed and
  dot-dashed lines in $N_{\rm H}$--$D$ frames show the empirical
  relations assuming $R=16$~km, 1~km and 20~km, respectively. See
  text for details. Vertical dashed lines in 1D distributions show 5\%, 50\% and 95\% quantiles.
  {\sl Bottom:} The 1D and 2D marginal posterior distributions for $N_{\rm H}$, $T$ and $D$
  in NSMAX+PL model without prior. Other options are the same as in the {\sl top} panel.
  The temperature $T$ is redshifted.
  }
\label{fig:ThermTriangle}
\end{figure*}

\begin{figure*}[t]
  \setlength{\unitlength}{1mm}
  \begin{picture}(160,190)(0,0)
    \put (0.0,95.0) {\includegraphics[scale=0.42]{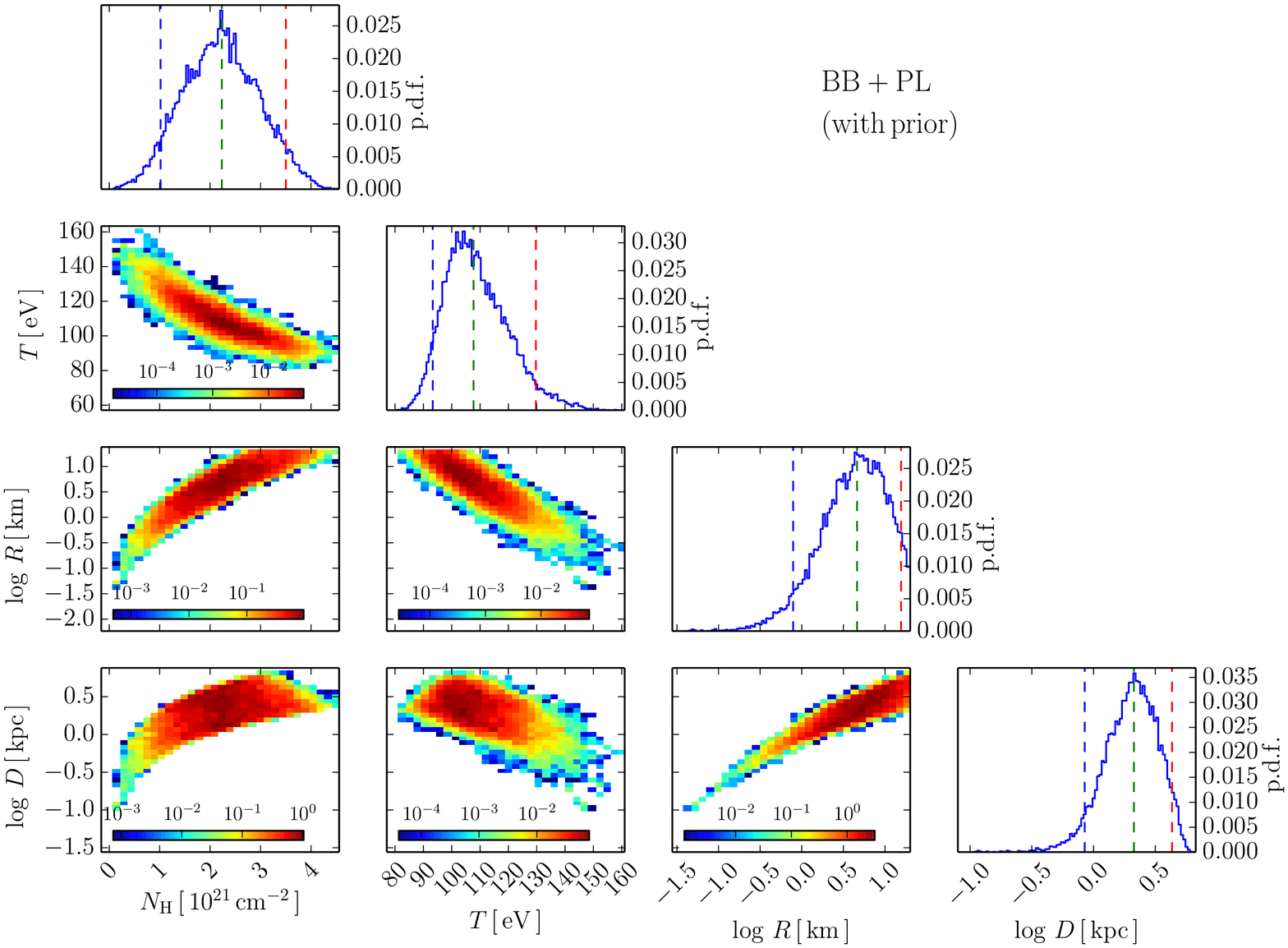}}
    \put (0.0,0.0) {\includegraphics[scale=0.42]{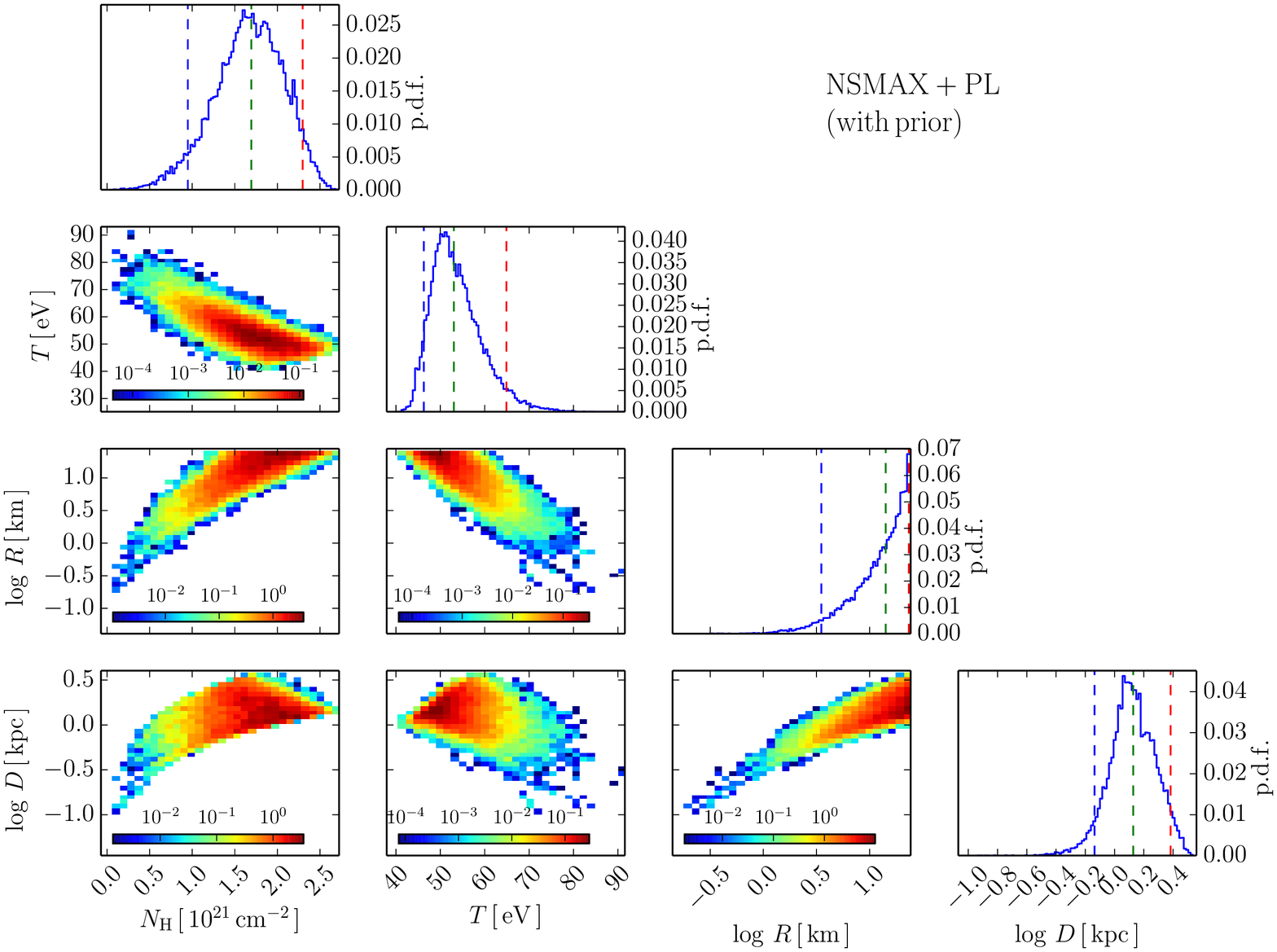}}
  \end{picture}
  \caption{1D and 2D marginal posterior distributions for $N_{\rm H}$, $T$, $R$ and $D$
  in the BB+PL model ({\sl top}) and  the NSMAX+PL model ({\sl bottom}), with account for the prior.
  Other options are the same as in
  Figure~\ref{fig:ThermTriangle}.
  }
\label{fig:ThermTrianglePrior}
\end{figure*}

It is instructive to consider which part of the posterior
parameters distribution can describe the physically allowed situation (in the selected
model framework). In particular, can the fit results for the thermal component
correspond
to the emission originating from the entire NS
surface or a part of it?
In Figure~\ref{fig:ThermTriangle}, we plot 1D and 2D
marginal posterior distributions for the parameters of BB and NSMAX models, which are
$N_{\rm H}$, temperature $T$ and the normalisation of the thermal
component. For both models, the temperature is given as measured by a distant observer.
The thermal component normalisation is presented in
the form of the distance to the pulsar if the apparent radius of
the emitting area is
$R$~= 16~km. As seen, the minimization of the
likelihood leads to correlation between $N_{\rm H}$ and normalisation of
the thermal component and thus to anti-correlation between $N_{\rm H}$
and distance $D$. In contrast, these latter quantities obviously
correlate in nature. It is unlikely to have a high $N_{\rm H}$ value at
a small distance and vice versa.

There exist empirical
models which provide dependence of interstellar absorption on distance.
Such models usually describe the correlation between distance
and optical extinction $A_V$. The latter can be transformed to $N_{\rm H}$
using one of the empirical relations between $A_V$ and $N_{\rm H}$,
for instance, the one given by \citet{predehl1995AsAp}.
To obtain the relation between
$A_V$ and distance in the pulsar direction,
we made use of the three-dimensional model of Galactic extinction
from \citet{drimmel2003AsAp} and also took into account the value of
maximal $N_{\rm H}$ in the pulsar direction,
$\sim$~(6--7)~$\times$~10$^{21}$ cm$^{-2}$ \citep{dickey1990ARAA,kalberla2005AsAp}.
This information can
be roughly summarized in a simple relation $N_{\rm H}$ [10$^{21}$
cm$^{-2}$]~$\approx$ $D$ [kpc] at $D<7$~kpc. We present this relation in the $D$--$N_{\rm H}$ plate in
Figure~\ref{fig:ThermTriangle}
with solid, dashed and dot-dashed lines
assuming $R$~= 16, 1 and 20 km,
respectively. The latter value is a reasonable theoretical
maximum for the NS apparent radius. Consider, for example, the BB
model. As can be seen, 16~km
radius is consistent with the data. The corresponding
$D$--$N_{\rm H}$ relation crosses the marginal posterior distribution  not far from
its maximum. The star in this case appears to be at about
$2.5$~kpc from the Sun and has the temperature of about 100~eV.
If the apparent radius is
$1$~km, our analysis leads (dashed line) to $N_{\rm H}\approx
10^{21}$~cm$^{-2}$ and hence places J0633 at $D\approx
1$~kpc. In this case, the inferred temperature should be higher, more than
120~eV. The radii much lower than 1 km would give worse fits and are unlikely.
The portion of the posterior distribution of parameters lying
down-right from the dot-dashed line in
Figure~\ref{fig:ThermTriangle} corresponds to unrealistic
$R>20$~km. This means that although the fit is formally good there,
the BB model with such parameters can not
describe thermal emission from the NS surface. Looking at
the $N_{\rm H}$--$T$ and $D$--$T$ plots we conclude that too low temperatures
are impossible, however the corresponding regions are broad and
this restriction is rather weak.
Similar analysis can be performed for the NSMAX model.
In this case, most of the
posterior distribution corresponds to the region with $R>20$~km.
However, radii of the order of $10$~km are still
allowed, giving $N_{\rm H}\approx 1.5\times10^{21}$~cm$^{-2}$ and
$D\approx 1.5$~kpc. This shrinks the possible temperature range
(in contrast to BB model). As clear from
Figure~\ref{fig:ThermTriangle},  temperature is constrained at $T\gtrsim
40$~eV and $N_{\rm H}\lesssim 2\times10^{21}$~cm$^{-2}$. Again, the low-temperature part of the posterior distribution
requires too high emitting area radii. Finally, as seen from
the plot, the $1$~km radius is too small if the NSMAX model is applied.

The Bayesian approach provides natural framework for inclusion
of the additional information,
such as the $D$--$N_{\rm H}$ relation discussed above,
by defining the appropriate prior distribution.
Moreover, it easily allows to take into account uncertainties in the prior knowledge.
Using this possibility, we incorporated the $D$--$N_{\rm H}$
relation as the Bayesian prior in the following way.
We made a conservative assumption
that the relation $N_{\rm H}$ [$10^{21}$ cm$^{-2}$]~$\approx$ $D$ [kpc]
is accurate up to a factor of two.
In principle, the central values
are more likely, and a bell-shaped form of the prior, for instance, Gaussian would be
appropriate. However, as the exact variances of the $D$--$N_{\rm H}$
relations are unavailable for us, we used the following flat prior:
0.5$D$ [kpc] $<$ $N_{\rm H}$ [$10^{21}$ cm$^{-2}$] $<$ 2$D$ [kpc]. In
addition, we constrained $D$ to be less than 7 kpc, which
is the approximate distance to the edge of the Galactic disk in the
pulsar direction. The application of such prior
naturally splits the thermal normalisation into two independent
parameters, $D$ and $R$. We also constrained the latter
to be less than 20~km.

The best-fit parameters inferred with account for the prior
are shown in third and fourth rows
of Table~\ref{t:x-fit}. Both BB and NSMAX models pass the
goodness-of-fit test.
The marginal 1D and 2D posterior distributions for $N_{\rm H}$, $T$, $D$
and $R$ are shown in Figure~\ref{fig:ThermTrianglePrior}.
We can see how the prior works, comparing
Figure~\ref{fig:ThermTrianglePrior} with Figure~\ref{fig:ThermTriangle}.
The distance is now
constrained by the prior relation and $N_{\rm H}$, and the constrains on $R$ are obtained
from $D$ and the thermal component normalisation.
The parameters in Table~\ref{t:x-fit} generally agree with the
qualitative considerations
presented above. We see that both BB and NSMAX models are consistent with
the physical picture where emission comes from the entire surface of the star.
For the BB model, however, the star should be somewhat more distant and more
absorbed than in case of the NSMAX model.
The hot-spot (of about $1$~km size) interpretation is also not
excluded under the BB model.
As it usually happens, the best-fit BB
temperature is about twice higher than that for a hydrogen atmosphere model \citep[e.g.,][]{2001pavlov}.

\section{DISCUSSION}
\label{sec:disc}

\subsection{Absorption feature}
\label{sec:disc_feature}

The analysis performed in Section~\ref{sec:an_feature}
favours presence of the absorption feature in J0633 spectra at
about 0.8~keV. Unfortunately, the shape of the feature is poorly
constrained with the current data, precluding us from plausible
interpretation of its nature. The first possibility to consider is the cyclotron line.
The cyclotron absorption line position, as seen by a distant observer,
for a particle of charge $Z$ and mass $m$ is given by
\begin{equation}
  \label{eq:cycl}
  E_{\rm cycl}^{\infty}=11.577 (1+z)^{-1} Z \frac{m_e}{m} B_{12}\
  {\rm keV},
\end{equation}
where $m_e$ is the electron mass and the line is assumed to form at the
NS surface. Now we can estimate a surface magnetic field as $B$~$\approx$
$8\times10^{10}$~G if the line is produced by electrons and
$B$~$\approx$ $1.4\times10^{14}$~G if it is produced by protons,
and even higher values for more massive ions. Both values are
inconsistent with the spin-down estimate of the dipole magnetic field,
$B$~= $4.9\times10^{12}$~G. The ``cyclotron'' field is much lower in  case of electrons and much
higher in case of protons.

Note that for other INS showing absorption lines, spin-down
magnetic fields, when determined, usually disagree with
``cyclotron''  magnetic fields. For SGR~0418+5729 and
PSR~J1740+1000, the discrepancy is as strong as for J0633. In
these cases, the proton cyclotron line interpretation is possible,
for instance, if there are strong small-scale (multipolar) surface
components of the magnetic field. The presence of the small-scale
fields is widely discussed in literature \citep[e.g.,][and
references therein]{Asseo2002,Harding2011}.
\citet{tiengo2013Natur} suggested this interpretation for the
spectral feature of SGR~0418+5729. They, however, had additional
arguments from the phase-resolved spectral analysis which is
unavailable in case of J0633. A somewhat similar feature was
recently detected in the spectrum of INS RX J0720.4$-$3125
\citep{borghese2015ApJ}. The feature is at $\sim$750 eV which
corresponds, if interpreted as the proton cyclotron line, to the
magnetic field about seven times higher than the spin-down one.

On the other hand, \citet{kargaltsev2012Sci} proposed that
the absorption feature in PSR~J1740+1000 is the electron cyclotron line which is
produced by a population of warm electrons occupying some regions
in the pulsar magnetosphere similar to Van Allen belts in the Earth
magnetosphere. If we adopt this interpretation for the line in J0633 and  assume that the magnetic field approximately
obeys the dipole law, $B\propto r^{-3}$, we can estimate the
position of the magnetospheric radiative belt as $r\approx 4 R_{\rm
NS}$, or about 30--40~km above the NS surface.

For the remaining objects, the disagreement is less dramatic.
The CCO 1E1207$-$5209 shows at least two features in the X-ray spectrum. Its low spin-down magnetic field
suggests the electron cyclotron interpretation. However, the position of the fundamental harmonic estimated
from the spin-down value is larger by a factor of 1.4
than the position of the strongest spectral feature \citep{gotthelf2013ApJ}.
Finally, for INSs from the magnificent seven group, spin-down magnetic fields are lower by a factor
of 1.1--7.2 than proton cyclotron magnetic fields estimates, with the best agreement achieved for
RX J1308.6+2127 \citep{pires2014AsAp}.

\begin{figure}[th]
\begin{center}
\setlength{\unitlength}{1mm}
 \includegraphics[scale=0.5]{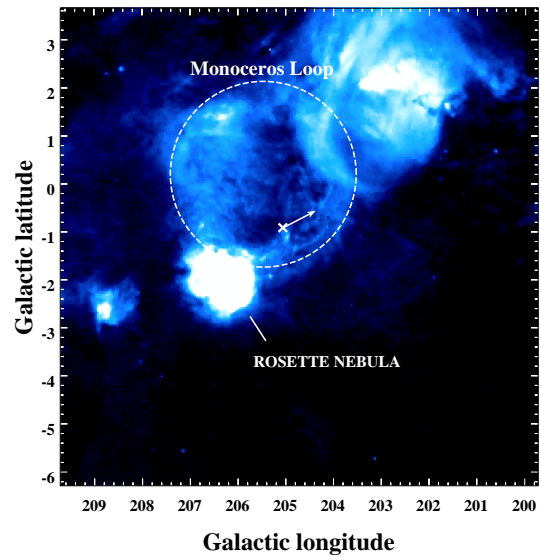}
   \caption{H$\alpha$ image of the Monoceros Loop
  region in Galactic coordinates taken from the Southern H-Alpha Sky Survey Atlas: H-Alpha \citep{Gaustard2001}.
  The Monoceros Loop SNR is marked by the dashed circle.
  The position of
  the pulsar and its possible proper motion direction are shown by $\times$ and
  the arrow, respectively. The Rosette nebula suggested as its likely birthplace is pointed on.
  }
  \label{fig:birth}
  \end{center}
\end{figure}

Another possible explanation of the absorption feature suggests
that it results from atomic transitions either in the stellar
atmosphere or in the interstellar medium. Varying abundances in
the model for the interstellar photoelectric absorption we found
that the feature could be explained assuming overabundance
of Fe along the pulsar line of sight. It is, in principle,
possible, as the J0633 position projects onto the Monoceros Loop
nebulosity, see Figure~\ref{fig:birth}, which was recognized as an
SNR from observations in radio, optical, and X-rays
\citep{Davies1963,Gebel1972,Davies1978,Leahy1985}. The distance to
the remnant is not exactly known, however most of the estimates
suggests a value of approximately $1.6$~kpc \citep[e.g.,][and
references therein]{BorcaJovanovic2009} providing an SNR shell
diameter of about $0.1$~kpc. This distance together with our
estimates for the J0633 distance (Table~\ref{t:x-fit}) allows the
pulsar to be behind the SNR and suffer from an additional
absorption of the SNR origin. However, to obtain a good fit for
the J0633 spectrum, too large Fe/H~$\approx$ $3\times10^{-4}$ is
required, which is about ten times as large as the solar one. To
provide such a high abundance of Fe on average along the pulsar
line of sight, abundance of Fe in the Monoceros Loop SNR itself
should be at least 100 times the solar abundance, which seems
implausible. Additionally, such a strong modification of the
interstellar absorption should also affect the PWN spectrum,
however, as stated above, we did not find any signature of
spectral features there. Finally, if the absorption feature is
formed in the mid-$Z$ atmosphere of NS, then generally broader and
weaker features are expected \citep{mori2007MNRAS}.

\subsection{J0633 in the view of the NS cooling theories}
\label{sec:disc_cool}

\begin{figure}[t]
  \begin{center}
   \includegraphics[scale=0.55]{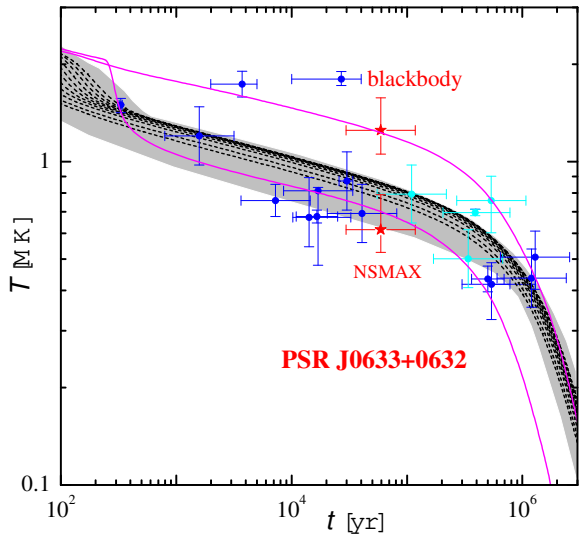}
  \end{center}
  \caption{Observations of isolated cooling NSs vs.
  cooling theory predictions.
  Temperatures obtained utilising the BB model are shown with the cyan colour,
  and those obtained with various atmospheric models are shown with the blue colour.
  The J0633 data points for BB and NSMAX models are shown by the star symbols.
  Dashed lines present the
  cooling curves corresponding to standard cooling of
  1--1.9$M_\odot$ NSs with the APR EOS. The filled
  region corresponds to a possible range of standard cooling
  curves including the unrealistically compact equations of state.
  Upper and lower solid curves illustrate the effects of the
  nuclear superfluidity in the NS core. Upper curve correspond to
  strong proton superfluidity suppression of the cooling, while
  lower curve is calculated including also Cooper pair formation
  emission from the triplet neutron superfluid.
  See text for details.}
\label{fig:cool}
\end{figure}

According to the Table~\ref{t:x-fit} and
Figure~\ref{fig:ThermTrianglePrior}, the J0633 thermal emission
can originate from the entire stellar surface. In this case, it
is instructive to compare the results with the NS cooling
theories. It is worth doing, according to Table~\ref{t:x-fit}, for
both BB and NSMAX models, although the inferred surface
temperatures are different. This comparison is performed in
Figure~\ref{fig:cool}, where the positions of J0633 on $T$--$\tau$ plane are
shown for both models along with
the data for other cooling isolated NSs.
The data on the latter objects are taken from the references collected
by \citet{yakovlev2008AIPC} and \citet{kaminker2009MNRAS},
excluding upper limits, with addition of several sources. The additional sources
include PSR J1741$-$2054 \citep{karpova2014ApJ},
PSR J0357+3205 \citep{kirichenko2014AsAp},
PSR J0007+7303 in the CTA~1 SNR \citep{caraveo2010ApJ,lin2010ApJ}
and two CCOs, where carbon atmosphere is used to describe their thermal radiation.
These are CCO in the Cas~A SNR \citep{yakovlev2011MNRAS},
the youngest source in Figure~\ref{fig:cool} and CCO XMMU J173203.3$-$344518 \citep{klochkov2015AsAp},
the hottest source in Figure~\ref{fig:cool}.
For the latter source we combined
temperatures for both distances given in \citet{klochkov2015AsAp}, and
multiplied the temperature errorbars by a factor of 2 to estimate 2$\sigma$
errors.
In Figure~\ref{fig:cool}, with cyan colour we show point obtained with
BB model, and with blue colour points obtained with various
atmospheric models, see the above cited references for details.
In Figure~\ref{fig:cool}, we also artificially adopt a factor of
two uncertainty on J0633 age.
It is seen, that if J0633 is covered
by the atmosphere, it is the one of the coldest middle-aged ($\tau \lesssim
10^5$~yr) isolated NS with measured surface temperature. If, in
contrast, the BB model actually fit the data, the inferred
temperature is much higher.

According to theory, isolated NSs cool down via the
neutrino emission from their interiors and via the photon emission
from their surfaces. The middle-aged stars are of particular
interest as they have isothermal interiors, except a thin heat
blanketing layer near the surface, and their cooling is dominated
by neutrino emission \citep[e.g.,][]{yakovlev2004ARA}.
Measurements of surface temperatures of such stars allow one to
determine the neutrino cooling rate and therefore to directly
explore the properties of matter deep inside the star.
By the filled region in Figure~\ref{fig:cool} we show the
predictions from the so-called standard NS cooling scenario which
assumes that a star has nucleon core and cools mainly via the
modified Urca processes of neutrino emission. According to
\citet{yakovlev2011MNRAS}, the standard cooling mainly depends on
the compactness of the star $x=R_{g}/R_{\rm NS}$, where $R_g$ is
the gravitational radius, or equivalently on the gravitational
redshift $1+z=(1-x)^{-1/2}$, being largely independent on the
particular NS model. More compact stars generally cool
faster. This property allows to estimate the neutrino cooling rate
of a particular middle-aged star without performing cooling
simulations. The filled region corresponds to a broadest possible region that
can be reached with the standard cooling (note that here the
standard iron heat-blanketing envelope is used). It includes also the
cooling curves corresponding to highly unrealistic equations of
state, which allow for extremely compact stars \citep[for
details see][]{yakovlev2011MNRAS}. For comparison, in Figure~\ref{fig:cool}
with short-dashed lines we present the cooling curves for stars
with a particular EOS in the core, that is the causal modification
\citep[same as used by, e.g.,][]{yakovlev2011MNRAS} of the  APR EOS
\citep{aprEOS}, widely used as a standard. The curves are given
for a range of NS masses from 1.0 to 1.9$M_\odot$ per
0.1 solar mass, plus the curve for the maximal mass $M_{\rm max}=1.929
M_\odot$ for this EOS. The direct Urca process is, in principle, allowed in the
massive stars with APR EOS, but here it is switched off. The
standard cooling curves for other reasonable EOSs basically fall
in  the same region. In other words, the part of the filled region
which is colder than the lowest dashed curve is reached in the
standard cooling scenario {\it in principle}, but is marginal.

Following the method described in \citet{yakovlev2011MNRAS}, we
found that the neutrino cooling rate of J0633 should be 30--1000
times stronger than the standard one, if the NSMAX model is applied, for a
reasonable star compactness $x<0.5$, and with account for a factor
of two uncertainty in the pulsar age. Only for unrealistically
compact stars, $x\approx 0.7$, the inferred NSMAX temperature can
be reached in the standard cooling models. A moderate increase in
the neutrino cooling rate $\lesssim 100$ can be   explained by
the minimal cooling theory which includes also the neutrino
emission in the process of Cooper pair formation in triplet
neutron superfluid in the NS core
\citep{page2004ApJS,gusakov2004AsAp}. We illustrate this
possibility with lower thin solid line in Figure~\ref{fig:cool}. It
corresponds to a 1.7$M_\odot$ APR EOS star and similar
superfluidity model as used by \citet{shternin2011MNRAS} in
explanation of the data on the NS in Cas~A SNR. Too lowering of temperature
is hardly possible in the minimal
cooling scenario. Nevertheless, in any case low temperatures of
cooling NSs can be explained if the direct Urca
processes are allowed in their cores. However, in order to get the
temperatures like J0633 has, assuming NSMAX spectral model, these
processes have to be suppressed, for instance, by superfluidity
\citep{yakovlev2004ARA}. Otherwise the enhancement of the neutrino
emission will be too strong.

For the BB model, the neutrino cooling rate, in contrast, should be
much weaker. We find that it must be suppressed by a factor of
10--300. The lowering by a factor $\lesssim 50$ is also possible
in the minimal cooling scenario if strong proton superfluidity is
involved which suppresses the conventional mUrca processes, and if
the internal temperature of the star is hotter than the neutron superfluidity
critical temperature so that Cooper pair emission does not operate
\citep{gusakov2004AsAp}. This is illustrated in
Figure~\ref{fig:cool} with the upper solid line which corresponds to
1$M_\odot$ star with the APR EOS and strong proton superfluidity in
the core. This curve fits nicely the BB data. Another possibility
allowed in the cooling theory is that the heat-blanketing envelope
of the star contains sufficient amount of the light elements. Then
the envelope is more transparent to heat and the star looks hotter
than the star with the same internal thermal state, but iron
(non-accreted) envelope \citep{Chabrier1997}. However, in the latter case the hydrogen (or
other light-element, for instance, carbon) atmosphere would be
more appropriate than BB to describe the emission spectra.
Also, the star will look hotter if magnetic field as strong as $\gtrsim10^{14}$~G
is present in the heat-blanketing envelope \citep{Potekhin2003}.
Finally, some additional heating mechanisms can operate in the
stellar interiors \citep[e.g,][]{yakovlev2004ARA}.

The inferred parameters of the BB model allow a different
interpretation of the NS thermal emission. It is possible that it
actually comes from a hot spot on a colder NS surface which is
heated by charged particles coming from the magnetosphere along
the magnetic field lines near the magnetic poles. The conventional
polar cap radius for J0633 can be estimated as $R_{\rm
cap}=0.145 (R_{\rm NS} [10^6~{\rm km}])^{3/2} (P [\rm s])^{-1/2}~{\rm
km}\approx 400$~m. This is inconsistent with the spectral fit
results (Table~\ref{t:x-fit}), however the 1--2~km emitting area
radii are possible. This range of radii correspond, according to
Figure~\ref{fig:ThermTrianglePrior}, to the temperatures $>125$ eV and
smallest possible distances of about 1--1.5~kpc. Note that these
values are favoured by the pseudo-distance relation, and also by
the possible birthplace of the pulsar in the Rosette nebula, see
below. In the hot spot picture, inferred temperature cannot be
compared with the cooling theories. The bulk of the surface is
then cold and invisible in X-rays.

\subsection{Non-thermal efficiencies and luminosities}
\label{sec:eff}

The distance ranges inferred from the  spectral fits (Table~\ref{t:x-fit})
allow to constrain the non-thermal luminosities of J0633
and its PWN. In Table~\ref{t:lum-eff} we give the 2--10~keV X-ray fluxes of the PWN, $F_{\rm X}^{\rm pwn}$ and the
non-thermal (PL) spectral component of the pulsar,
$F_{\rm X}^{\rm psr}$, for both spectral models used\footnote{Two models with prior from
two last rows in Table~\ref{t:x-fit}.}.
As expected, these values almost do not depend on the type  of the thermal continuum model (BB or NSMAX).
Corresponding non-thermal luminosities $L_{\rm X}^{\rm pwn}$ and $L_{\rm X}^{\rm psr}$
are also given in Table~\ref{t:lum-eff}
along with the values of efficiencies $\eta_{\rm X}^{\rm psr}=L_{\rm X}^{\rm psr}/\dot{E}$ and
$\eta_{\rm X}^{\rm pwn}=L_{\rm X}^{\rm pwn}/\dot{E}$.
These values show some dependence on a choice of the spectral model because in the two models the
inferred distance ranges are slightly different (Table~\ref{t:x-fit}).
In any case, the parameters of the X-ray non-thermal emission in Table~\ref{t:lum-eff} are
not peculiar in comparison with those
for other pulsars with similar $\dot{E}$ \citep{Kargaltsev2008, danilenko2013AsAp}.
In addition, in Table~\ref{t:lum-eff}, we show J0633
$\gamma$-ray luminosities $L_\gamma^{\rm psr}$ and corresponding efficiencies
$\eta_\gamma^{\rm psr}= L_\gamma^{\rm psr}/\dot{E}$.
They are calculated basing on the pulsar $\gamma$-ray flux
$F_{\rm \gamma}^{\rm psr}$~= $(9.4\pm 0.5)\times10^{-11}$ erg~cm$^{-2}$~s$^{-1}$
\citep{abdo2013ApJS}.
Note, that for large distances,
$D$~$\gtrsim$~3 kpc, allowed for the BB+PL model (Table~\ref{t:lum-eff}),
the $\gamma$-ray efficiency is higher than 1.
However, there are $\gamma$-ray pulsars with precisely measured distances which have
$\eta_{\rm \gamma}>1$ \citep{abdo2013ApJS}.
The apparent violation of the energy conservation law is usually resolved
by account for unknown beaming of the $\gamma$-ray emission.

\begin{table*}[th]
\scriptsize
 \caption{Non-thermal fluxes, luminosities and efficiencies.
  }\label{t:lum-eff}
  \begin{tabular}{l*{9}{c}}
\\ \hline\multicolumn{1}{c}{} \\
Model    & $\log F_{\rm X}^{\rm psr}$ & $\log L_{\rm X}^{\rm psr}$ & $\log \eta_{\rm X}^{\rm psr}$ & $\log F_{\rm X}^{\rm pwn}$ & $\log L_{\rm X}^{\rm pwn}$ & $\log \eta_{\rm X}^{\rm pwn}$ & $\log L_{\rm \gamma}^{\rm psr}$ & $\log \eta_{\rm \gamma}^{\rm psr}$ \\
         & (erg~cm$^{-2}$~s$^{-1}$)  & (erg~s$^{-1}$)            &                               & (erg~cm$^{-2}$~s$^{-1}$) & (erg~s$^{-1}$)            &                               & (erg~s$^{-1}$)                &     \\
\hline
\multicolumn{7}{c}{}\\
BB+PL    & $-13.4^{+0.2}_{-0.2}$  & $31.4^{+0.6}_{-0.8}$       &  $-3.7^{+0.6}_{-0.8}$         & $-12.6^{+0.1}_{-0.1}$ & $32.1^{+0.6}_{-0.8}$     & $-3.0^{+0.6}_{-0.8}$       & $34.7^{+0.6}_{-0.8}$      & $-0.4^{+0.6}_{-0.8}$ \\
\multicolumn{7}{c}{}\\
NSMAX+PL & $-13.3^{+0.2}_{-0.2}$  &  $31.0^{+0.5}_{-0.6}$       & $-4.1^{+0.5}_{-0.6}$          & $-12.6^{+0.1}_{-0.1}$ & $31.7^{+0.5}_{-0.5}$     & $-3.4^{+0.5}_{-0.5}$       & $34.3^{+0.5}_{-0.5}$ & $-0.8^{+0.5}_{-0.5}$ \\

\multicolumn{7}{c}{}\\
\hline
\end{tabular}
\tabnote{X-ray fluxes, luminosities and efficiencies are calculated
in the 2--10 keV range.}
\end{table*}

\subsection{Presumed birthplace}
\label{sec:birth}

The J0633 PWN morphology
and extent  (Figure~\ref{fig:chandra}) is reminiscent of, for instance, a well-studied
bowshock ``Mouse'' nebula (G359.23$-$082) powered by the
fast-moving PSR~J1747$-$2958  \citep[e.g.,][]{hales2009ApJ}. The
similarity with the Mouse suggests the direction of a proper
motion of J0633 as shown by the long arrow in
Figure~\ref{fig:chandra}. Adopting a typical synchrotron cooling
time of X-ray emitting electrons in PWNe of $\sim$~1000 yr
\citep[e.g.,][]{kargaltsev2008ApJ} and the J0633 PWN tail size of
1\farcm3 (see Figure~\ref{fig:chandra}), we estimated the pulsar
proper motion as 80 mas~yr$^{-1}$, suggesting the pulsar angular
displacement of 1\fdg3 during its lifetime of $\sim$~60~kyr. Looking
backwards  the assumed proper motion
direction on the
extended field of view  (Figure~\ref{fig:birth}),  we found a likely
birthplace of the pulsar at the estimated angular displacement --
the Rosette nebula. It is known as a young, 50 Myr, active star
forming region located at the edge of
the Monoceros Loop SNR. The   distance to the Rosette of $\sim$~1.5~kpc \citep{Ogura1981} is
compatible with the J0633 distance estimates, discussed above. Looking from the other side,
adopting the Rosette nebula as a likely birthplace we can
independently estimate the distance to the pulsar. Taking into
account the conservative  uncertainties of a factor of two for the
pulsar age and angular distance to its specific birth location
inside the Rosette nebula, and assuming that the pulsar 3D
velocity is smaller than 2000 km~s$^{-1}$
\citep[e.g.][]{Hobbs2005}, we obtain a distance range of $1.2<D<1.8$~kpc adopting the
distance to the Rosette nebula of 1.4--1.6~kpc. This range is consistent with that derived from the spectral analysis and
does not put additional constraints on the thermal
emission models.

\section{CONCLUSIONS}
\label{sec:concl}

We have analysed in detail the X-ray spectrum of the
$\gamma$-ray pulsar J0633+0632. We found the evidence of the narrow
absorption feature in its spectrum at $804^{+42}_{-26}$~eV
with the equivalent width of $63^{+47}_{-36}$ eV,
where errors are at 90\% confidence.
While the shape of the feature cannot be constrained with the current data, the failure of
any smooth continuum model in vicinity of $0.8$~keV is statistically proven.
We briefly discussed possible physical
interpretations of the detected feature and currently favoured the cyclotron nature.

Apart from the spectral feature, we investigated the
properties of the X-ray continuum. We confirmed the conclusion by
\citet{ray2011ApJS} that the soft part of the J0633 spectrum is
dominated by the thermal component, and the hard tail is described
by the non-thermal PL. We went further and took into account the
correlation between the distance to the pulsar and the
interstellar absorption along the J0633 line of sight on the basis
of the empirical distance-extinction maps. It was included in the
form of the prior distribution for the model parameters. In
addition, the PWN spectrum was fitted simultaneously with the
pulsar spectrum, allowing to better constrain the $N_{\rm H}$
parameter. As a result of this analysis, we found that the thermal
emission possibly originates from the entire surface of the star
and its spectrum can be equally well described by the blackbody or
the magnetised hydrogen atmosphere models. In the BB case, the hot
spot origin of the thermal emission is also possible. The distance
to the pulsar was constrained within 1--4~kpc range basing on the
X-ray spectral analysis. This is especially important, as the
dispersion measure distance is unavailable for the radio-silent
J0633.

Confronting the inferred temperatures with the data on other
cooling NSs and predictions from NS cooling theories, we found that for
the atmospheric spectral model, J0633 is one of the coldest middle-aged
NS with measured temperature. In this case it must cool considerably faster than
the standard cooling scenarios predict.
In contrast, if the spectrum of the substantial part of the
NS surface is blackbody, then J0633 is hotter than a standard cooling
NS.

In addition, we found a possible birth
site of the pulsar -- the Rosette nebula. This, along with the
shape of the PWN, suggests that J0633 can have prominent
proper motion. At the moment, these findings do not impose
additional constrains on the pulsar properties.
When this paper was in preparation,
deeper observations of J0633 with XMM-Newton were approved for the
AO-14 observational cycle. 
If these observations are performed, more elaborate consideration of the 
absorption feature origin and nature of the continuum emission 
will be possible.
The detection of X-ray pulsations 
will be especially useful to attribute the thermal emission to  
a small hot region or to the entire surface of the NS. 
The study of the variation of the absorption feature with the rotational
phase is also important
to explore its origin and, if the feature is a cyclotron line, magnetic field geometry \citep{kargaltsev2012Sci,tiengo2013Natur,borghese2015ApJ}.
The observations will also allow to find out which  of the two models, 
blackbody or hydrogen atmosphere, is more appropriate to describe the pulsar spectrum.
If the future data will favour the atmosphere model and will confirm 
that the surface temperature is as low as it follows from the current data, then it will result 
in important consequences for understanding the physics of the neutrino emission in dense cores of NSs.


\begin{acknowledgements}
The scientific results reported in this article
are based on data obtained from the Chandra Data Archive,
observations made by the Chandra X-ray Observatory. We thank Dima
Barsukov and Serge Balashev for helpful discussion.
YS acknowledges support from the Russian Foundation for
Basic Research (grants 13-02-12017-ofi-m, 14-02-00868-a). The work of AD, DZ and AK was supported by RF
Presidential Programme MK-2837.2014.2.
The modelling of the NS cooling scenarios in Sec. 3.2 was performed by PS under the
support of the Russian Science Foundation, project 14-12-00316.
\end{acknowledgements}

\bibliographystyle{apj}
\bibliography{ref}

\end{document}